\documentclass[twocolumn,prl,letter]{revtex4}
\usepackage{amssymb}
\usepackage{graphicx}
\usepackage{natbib}

\begin{document}

\title{Magnetic fullerenes inside single-wall carbon nanotubes}
\author{F. Simon,$^{1,3}$ H. Kuzmany,$^{1}$ B. N\'{a}fr\'{a}di,$^{2}$ T. Feh%
\'{e}r,$^{2,3}$ L. Forr\'{o},$^{2}$ F. F\"{u}l\"{o}p,$^{3}$ A. J\'{a}nossy,$%
^{3}$, L. Korecz,$^{4}$ A. Rockenbauer,$^{4}$ F. Hauke,$^{5}$ and A.
Hirsch$^{5}$} \affiliation{$^{1}$Institut f\"{u}r Materialphysik,
Universit\"{a}t Wien, Strudlhofgasse 4, A-1090 Wien, Austria}
\affiliation{$^{2}$Institute of Physics of Complex Matter, FBS Swiss
Federal Institute of Technology (EPFL), CH-1015 Lausanne,
Switzerland} \affiliation{$^{3}$Budapest University of Technology
and Economics, Institute of Physics and Solids in Magnetic Fields
Research Group of the Hungarian Academy of Sciences, H-1521,
Budapest P.O.Box 91, Hungary} \affiliation{$^{4}$Chemical Research
Center, Institute of Chemistry, P.O.Box 17, H-1525 Budapest,
Hungary}
\affiliation{$^{5}$Institut f\"{u}r Organische Chemie der Friedrich Alexander Universit%
\"{a}t Erlangen-N\"{u}rnberg, Henkestrasse 42, D - 91054 Erlangen}

\begin{abstract}
C$_{59}$N magnetic fullerenes were formed inside single-wall carbon
nanotubes by vacuum annealing functionalized C$_{59}$N molecules
encapsulated inside the tubes. A hindered, anisotropic rotation of
C$_{59}$N was deduced from the temperature dependence of the
electron spin resonance spectra near room temperature. Shortening of
spin-lattice relaxation time, $T_{1}$, of C$_{59}$N indicates a
reversible charge transfer toward the host nanotubes
above $\sim 350$ K. Bound C$_{59}$%
N-C$_{60}$ heterodimers are formed at lower temperatures when
C$_{60}$ is co-encapsulated with the functionalized C$_{59}$N. In
the 10-300 K range, $T_{1}$ of the heterodimer shows a relaxation
dominated by the conduction electrons on the nanotubes.
\end{abstract}
\maketitle



Single-wall carbon nanotubes (SWCNTs)
\cite{IijimaNAT1993,BethuneNAT1993} exhibit a variety of unusual
physical phenomena related to their one-dimensional and strongly
correlated electronic properties. These include excitonic effects
\cite{LouiePRL2004,HeinzSCI2005}, superconductivity
\cite{TangSCSCI}, the Tomonaga-Luttinger liquid state
\cite{KatauraNAT2003}, and the Peierls transition
\cite{BohnenPeierlsPRL2004}. Magnetic resonance is a powerful method
to study strong correlations in low dimensional systems. However,
for SWCNTs both nuclear magnetic resonance (NMR) and electron spin
resonance (ESR) are severely limited by NMR active $^{13}$C nuclei
and ESR active electron spins in residual magnetic catalytic
particles and other carbon phases. Synthesis of $^{13}$C isotope
engineered SWCNTs solved the problem for NMR
\cite{SimonPRL2005,SingerPRL2005}. To enable ESR spectroscopy of
SWCNTs, a local probe, specifically attached to SWCNTs, is required. The N@C$_{60}$ \cite%
{WeidingerPRL} and C$_{59}$N \cite{WudlReview} magnetic fullerenes
are ideal candidates for such studies. In fullerene doped SWCNTs,
fullerenes occupy
preferentially the interior of the tubes and form "peapods" (C$_{60}$@SWCNT) \cite%
{SmithNAT}. Fullerenes adhesing to the outside can be removed
\cite{KatauraSM2001} in contrast to e.g. filling with iron
\cite{LuzziJNN2003}. ESR on encapsulated magnetic fullerenes could
yield information on the electronic state of the tubes and it could
also enable to study the fullerene rotational dynamics in a confined
environment. In addition, magnetic fullerene peapods could exploit
the combination of the SWCNT strength and the magnetic moment of
molecules in magnetic scanning probe tips and they could enable a
bottom-up design for magnetic storage devices or for building
elements of quantum computers \cite{HarneitPSS}.

Typical spin concentrations in (N@C$_{60}$:C$_{60}$)@SWCNT are low,
$\sim $1 spin/tube, and the N spins are insensitive to SWCNT
properties \cite{SimonCPL2004}. The C$_{59}$N monomer radical is a
better local probe candidate as the unpaired electron is on the
cage. C$_{59}$N can be chemically prepared but it forms spinless
dimers (C$_{59}$N)$_{2}$ or monomer adducts \cite{WudlReview}. The
magnetic C$_{59}$N monomer radical can be stabilized as
C$_{59}$N:C$_{60}$, a dilute solid solution of C$_{59}$N in C$_{60}$
\cite{FulopCPL}.

Here, we report on the first ESR study of SWCNT properties and
peapod rotational dynamics using a paramagnetic local probe:
C$_{59}$N monomer radicals encapsulated inside SWCNTs. SWCNTs were
first filled with chemically inert C$_{59}$N derivatives. A heat
treatment in vacuum removes the side-group and the monomer radical
is left behind. The rotation of encapsulated C$_{59}$N is hindered
and anisotropic in contrast to the isotropic rotation in
C$_{59}$N:C$_{60}$. In samples with co-encapsulated C$_{60}$ and
C$_{59}$N, bound C$_{59}$N-C$_{60}$ heterodimers are formed during
the heat treatment. The electron spin-relaxation time of the
heterodimer is dominated by the conduction electrons of the SWCNTs
and follows the Korringa law.

SWCNTs were filled with air stable C$_{59}$N derivatives
(4-Hydroxy-3,5-dimethyl-phenyl-hydroazafullerene, C$_{59}$N-der in
the following) and C$_{59}$N-der:C$_{60}$ in concentrations of 1:10.
The mean value of the SWCNT
diameter distribution, as determined from Raman studies \cite{KuzmanyEPJB}, $%
d$ = 1.40 nm is optimal for fullerene encapsulation. A mixture of
dissolved fullerenes and SWCNTs were sonicated in toluene and
filtered, which results in a high degree of encapsulation as shown
by transmission electron microscopy and Raman spectroscopy in Ref.
\cite{SimonCar2006}. The peapods were
mixed with ESR silent SnO$_{2}$ to separate the conducting SWCNT pieces and were annealed in dynamic vacuum at 600 $%
^{\circ }$C for 15 minutes to remove the side-group. The
air-sensitive materials were sealed under He in quartz tubes. ESR
was studied on a Bruker Elexsys spectrometer at 9 GHz in the 10-600 K temperature range with spin-sensitivity calibrated by CuSO$_{4}\cdot $%
5(H$_{2}$O$)$.

\begin{figure}[tbp]
\includegraphics[width=0.75\hsize]{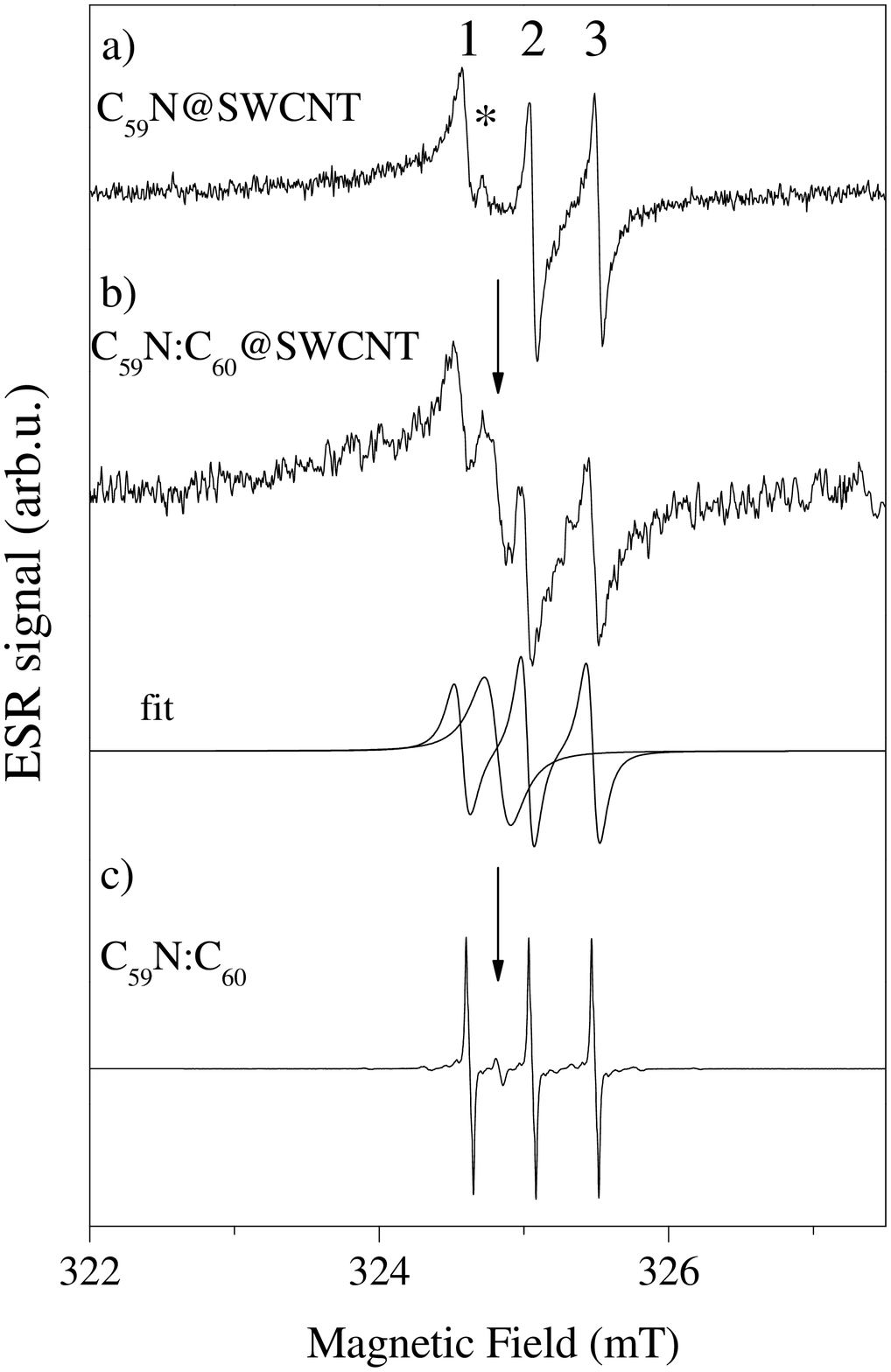}
\caption{ESR spectra of a) C$_{59}$N@SWCNT, b)
C$_{59}$N:C$_{60}$@SWCNT, and c) crystalline C$_{59}$N:C$_{60}$ at
300 K. 1, 2, and 3 denote the $^{14}$N hyperfine triplet lines with
nuclear state of $I=1$, 0, and -1, respectively. Asterisk in a)\
shows
a weak impurity signal. Arrows in b) and c) indicate the C$_{59}$N-C%
$_{60}$ heterodimer. Fit in b) shows the deconvolution of the ESR
signal into the $^{14}$N triplet of monomer C$_{59}$N and the
C$_{59}$N-C$_{60}$ heterodimer components.} \label{ESRspectra}
\end{figure}

Fig. \ref{ESRspectra} shows the room temperature ESR spectra of
C$_{59}$N@SWCNT (sample A) and C$_{59}$N:C$_{60}$@SWCNT (sample B)
and for comparison the spectrum of crystalline C$_{59}$N:C$_{60}$
(sample C) from Ref. \cite{FulopCPL}. This latter spectrum was
previously assigned to the superposition of rotating C$_{59}$N
monomers and bound C$_{59}$N-C$_{60}$ heterodimers
\cite{RockenbauerPRL2005}. The large spin density at the $^{14}$N
nucleus of the rotating C$_{59}$N molecule results in an ESR triplet
signal and the C$_{59}$N-C$_{60}$ heterodimer has a singlet signal
(arrow in Fig. \ref{ESRspectra}c) as the spin density resides on the
C$_{60}$ molecule. $^{14}$N triplet structures are observed in the
peapod samples (A and B) with identical hyperfine coupling as in the
crystalline sample (C) and are thus identified as the ESR signals of
rotating C$_{59}$N monomer radicals encapsulated inside SWCNTs. The
additional component (arrow in Fig. \ref{ESRspectra}b) observed for
sample B, which contains co-encapsulated C$_{60}$, is identified as C$_{59}$%
N-C$_{60}$ heterodimers encapsulated inside SWCNTs since this signal
has the same $g$-factor as in the crystalline material. This singlet
line is absent in sample A which does not contain C$_{60}$. For both
peapod samples a broader line with HWHM of $\Delta H \sim 0.6$ mT is
also observed. The broader component appears also on heat treatment
of reference samples without encapsulated C$_{59}$N-der and is
identified as a side-product. Annealing at 600 $^{\circ }$C is
optimal: lower temperatures result in smaller C$_{59}$N signals and
higher temperatures increase the broad impurity signal without
increasing the C$_{59}$N intensity.

Deconvolution of the ESR signal (Fig. \ref{ESRspectra}b) and the
intensity calibration against the CuSO$_{4}\cdot $%
5(H$_{2}$O$)$ spin standard allows to measure the amount of
C$_{59}$N related (C$_{59}$N monomer and C$_{59}$N-C$_{60}$
heterodimer) spins in the sample. The amount of encapsulated
fullerenes is known \cite{SimonPRL2005}, thus the ratio, $r$, of
observed C$_{59}$N related spins and encapsulated C$_{59}$N-der can
be determined. We obtained $r=2.5(6)$ \% and $r=12(3)$ \% for the A
and B samples, respectively. The observed ESR signal of C$_{59}$N
can be reduced due to various reasons: i) dimerization of C$_{59}$N
first neighbors into ESR silent (C$_{59}$N)$_{2}$, ii) incomplete
transformation of C$_{59}$N-der into C$_{59}$N, iii) dipolar fields
of near neighbor C$_{59}$N pairs: only those C$_{59}$N related spins
are observed which have no neighbors with dipole fields larger than
the ESR line-width, $\sim$ 0.07 mT. The likely origin of the smaller
$r$ value for sample A is dimerization. For sample B, however, a
statistical calculation of the dipolar fields gives $r=9$ \% in
agreement with the experimental value. For the calculation, the
fullerene lattice constant inside the tubes \cite{HiraharaPRB}, the
three dimensional arrangement of the SWCNTs into bundles
\cite{DresselhausTubesNew}, the random orientation of the bundles,
and the concentration of C$_{59}$N was taken into account. As a
result, the data for sample B supports that most C$_{59}$N-der is
transformed to C$_{59}$N monomer radicals.

\begin{figure}[tbp]
\includegraphics[width=0.75\hsize]{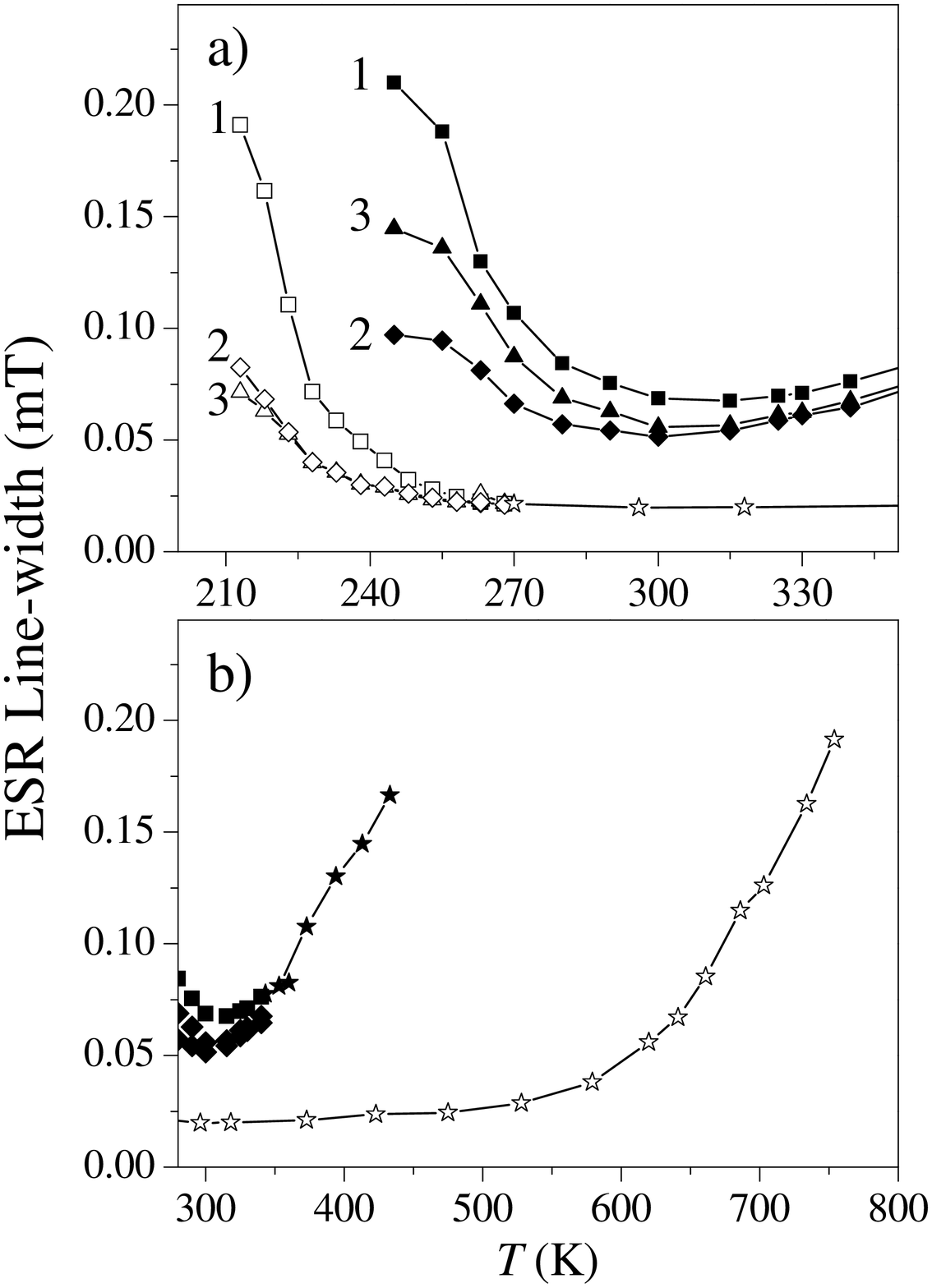}
\caption{Line-widths of the $^{14}$N triplet components in
crystalline C$_{59}$N:C$_{60}$ (open symbols) and (C$_{59}$N,
C$_{59}$N:C$_{60}$)@SWCNT (full symbols) samples below 350 K (a) and
above 290 K (b). Stars show the data for the temperature range where
the line-widths of the three components are equal. Solid lines are
guides to the eye.} \label{widths}
\end{figure}

The temperature dependence of the line-widths of the $^{14}$N
triplet is identical for the two peapod samples, A and B, and is
shown together with the data on C$_{59}$N:C$_{60}$ in Fig.
\ref{widths} using the labeling given in Fig. \ref{ESRspectra}a. The
line-widths are $\sim$ 0.04 mT larger for the peapod than for the
crystalline material. This excess line-width is similar to that of
(N@C$_{60}$:C$_{60}$)@SWCNT and is related to the stray magnetic
field of magnetic catalytic particles in the nanotube sample
\cite{SimonCPL2004}. The three C$_{59}$N triplet lines are broadened
unequally at lower temperatures for both the peapod and crystalline
materials. The details of the low temperature broadening are
different for the two kinds of materials: for encapsulated
C$_{59}$N, the unequality persists to higher temperatures and the
three lines broaden differently, whereas for the crystalline
C$_{59}$N:C$_{60}$ line 1 broadens significantly and lines 2 and 3
broaden equally but less.

The unequal broadening of $^{14}$N triplet lines with decreasing
temperature is well known for NO spin labels and is explained by an
incomplete motional narrowing of the anisotropic hyperfine and
$g$-factor anisotropy \cite{FreedBook}. For crystalline
C$_{59}$N:C$_{60}$, molecular rotation becomes rapid enough
immediately above the 261 K structural phase transition to result in
motionally narrowed lines \cite{FulopCPL}. In contrast, the
line-width data of encapsulated C$_{59}$N indicates a hindered
rotation. The line-width in the hindered molecular rotation regime
is:
\begin{equation}
\label{broadening} \Delta H =A+BM_{I}+CM_{I}^{2}
\end{equation}

\noindent where $M_{I}$ is the nuclear state of the $^{14}$N
hyperfine lines and the parameters $A$, $B$, and $C$ depend on the
hyperfine and $g$-tensor components, on the $R_{x}$, $R_{y}$, and
$R_{z}$ molecular rotational rates around each axis, and on the ESR
frequency \cite{FreedBook}. For crystalline C$_{59}$N:C$_{60}$ at 9
GHz the isotropic rotation of the molecule (i.e.
$R_{x}=R_{y}=R_{z}$) combined with the hyperfine and $g$-tensors
results in $B \approx C$ and thus in an equal broadening for lines 2
($M_{I}=0$) and 3 ($M_{I}=-1$) \cite{FulopCPL,RockenbauerPRL2005}.
The unequality of the line-widths for encapsulated C$_{59}$N
indicates an anisotropic rotation, i.e. $R_{x}\neq R_{y} \neq R_{z}$
as the hyperfine and $g$-tensor components are expected to be
identical for the peapod and crystalline materials. The anisotropic
rotation is suggested to originate from the anisotropic environment
inside the nanotubes.

Above 300 K the triplet line-widths rapidly grow with temperature
(Fig. \ref{widths}b) and the signal intensity decreases faster than
the Curie-law but no new lines appear. The broadening and the loss
of signal intensity is fully reversible with temperature cycling.
The broadening is reminiscent of that observed above $\sim$ 600 K in
the crystalline material, which was interpreted as a spin-lattice
life-time
shortening due to delocalization of the electron from C$_{59}$N over the C$%
_{60}$ matrix \cite{RockenbauerPRL2005}. Based on the analogous
behavior, we suggest that reversible charge transfer from C$_{59}$N
toward the nanotubes takes place above $\sim$ 350 K. The
significantly lower temperature of the charge transfer indicates a
larger overlap of the extra electron of C$_{59}$N to the SWCNTs
compared to its overlap with the C$_{60}$
conduction band in the crystalline material. In C$_{59}$N:C$%
_{60}$ the broadening is accompanied by the emergence of the ESR
signal of the delocalized electrons. The intrinsic ESR signal of
SWCNTs is not observable \cite{NemesPRB2000,SalvetatPRB2005}, which
explains the absence of a signal corresponding to charge transferred
electrons on the tubes.

\begin{figure}[tbp]
\includegraphics[width=0.8\hsize]{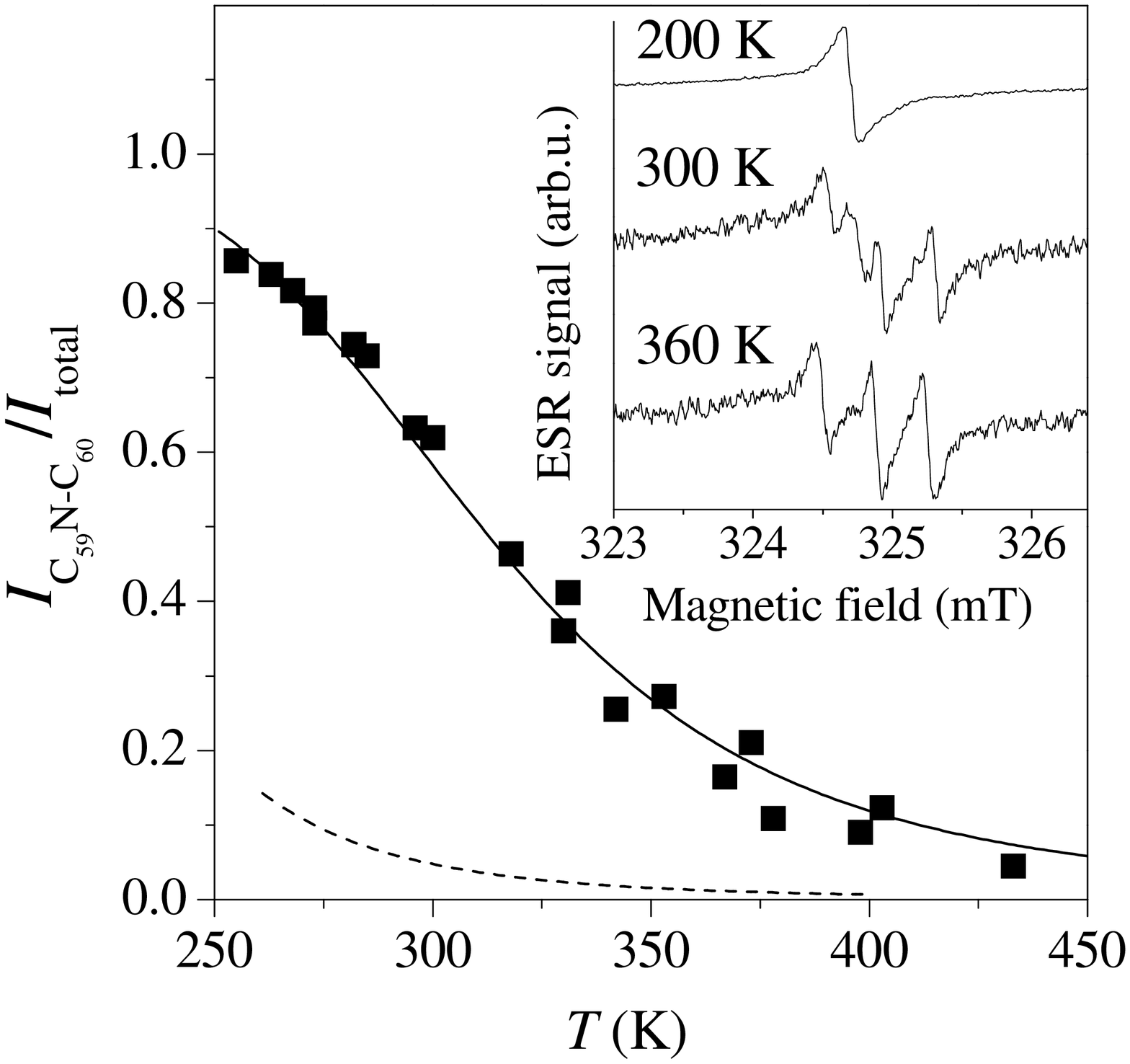}
\caption{Concentration of C$_{59}$N-C$_{60}$ bound heterodimers in C$_{59}$%
N:C$_{60}$@SWCNT. Solid curve is a fit with parameters explained in
the text. Dashed curve shows the same quantity for crystalline
C$_{59}$N:C$_{60}$ above the 261 K phase transition. Note the much
higher heterodimer concentration for the peapod material. Inset
shows the temperature evolution of the spectra.} \label{heterodimer}
\end{figure}

The coexistence of bound C$_{59}$N-C$_{60}$ heterodimers and
rotating C$_{59} $N molecules was understood for C$_{59}$N:C$_{60}$
as a thermal equilibrium between the ground state heterodimer and
the rotating monomers \cite{RockenbauerPRL2005}. The inset in Fig. \ref{heterodimer}
shows a similar behavior for C$_{59}$%
N:C$_{60}$@SWCNT: the heterodimer dominates the low temperature
spectrum and vanishes at higher temperatures, however the relative
intensity of the heterodimer is much larger in this material. The
heterodimer signal intensity normalized by the total
(heterodimer+triplet) intensity gives the heterodimer concentration and is shown in Fig. \ref%
{heterodimer}. Similarly to crystalline C$_{59}$N:C$_{60}$, the
heterodimer concentration can be fitted with:


\begin{eqnarray}
\frac{I_{\text{C}_{59}{\text{N}-\text{C}_{60}}}}{I_{\text{total}}}=\frac{1}{\left(
1+\text{e}^{ (-E_{\text{a}}/T+\Delta S)}\right)}
\label{Heterodimer_Intensity}
\end{eqnarray}

\noindent where $E_{\text{a}}$ is the binding energy of the
heterodimer and $\Delta S$ is the entropy difference between the
rotating monomer and the static heterodimer states. A fit with Eq.
\ref{Heterodimer_Intensity} for the peapod material is
shown in Fig. \ref{heterodimer} as a solid curve and gives $E_{\text{a}}(\text{peapod})=2800(200)$ K and $%
\Delta S(\text{peapod})=9(1)$. This compares to the results for the
crystalline material with $E_{\text{a}}(\text{cryst})=2400(600)$ K
and $\Delta S(\text{cryst})=11(2)$ \cite{RockenbauerPRL2005}. The
higher heterodimer concentrations for the peapod material is caused
by the larger $E_{\text{a}}$ and smaller $\Delta S$ values. The
latter is explained by the limited rotational freedom of encapsulated C$%
_{59}$N.

\begin{figure}[tbp]
\includegraphics[width=0.8\hsize]{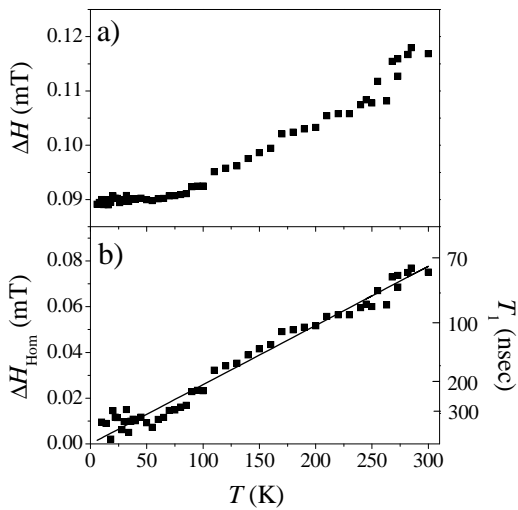}
\caption{a) Temperature dependence of the ESR line-width for the
heterodimer signal. b) The homogeneous contribution to the
line-width with a linear fit (solid line). The corresponding $T_{1}$
values are shown on the right axis.} \label{linewidth}
\end{figure}

Similar to NMR spectroscopy, the ESR spin-lattice relaxation time,
$T_{1}$, of the localized heterodimer spins yields information on
the electronic structure of the host SWCNT material
\cite{SlichterBook}. $T_{1}$ can be measured by time-resolved ESR
measurements or in continuous wave ESR spectroscopy from the
line-width, $\Delta H$, by separating the homogeneous, relaxation
related line-width from the inhomogeneous one. In Fig.
\ref{linewidth}a, we show $\Delta H$ for the heterodimer signal as
determined from fits with derivative Lorentzian lines. Clearly,
$\Delta H$ has a temperature dependent component in addition to a
$\Delta H_{0}=0.089(2)$ mT residual line-width, which is obtained by
averaging the line-widths below 50 K. The line-shape of the
heterodimer signal does not change with temperature, which indicates
a uniform, homogeneous broadening in addition to the inhomogeneous
residual width.

To obtain the homogeneous line-width, $\Delta H_{\text{Hom}}$, we
subtracted $\Delta H_{0}$ from the line-width data: $\Delta
H_{\text{Hom}}=\sqrt{\Delta H^{2}-\Delta H_{0}^{2}}$. Fig.
\ref{linewidth}b shows $\Delta H_{\text{Hom}}$ and $%
1/T_{1}=\gamma_{\text{e}} \Delta H_{\text{Hom}}$, where
$\gamma_{\text{e}}/2 \pi=28.0 \text{ GHz/T}$ is the electron
gyromagnetic ratio. $1/T_{1}$ as a function of $T$ is linear, with
$(T_{1}T)^{-1}= 4.2(2)\cdot 10^{4} (\text{sK})^{-1}$ (fit shown in
Fig. \ref{linewidth}b), which suggests that Korringa relaxation,
i.e. the interaction with conduction electrons \cite{SlichterBook}
gives the relaxation of the heterodimer. An effective coupling
constant (averaged for tube chiralities), $A$, of localized spins
and conduction electrons is 11 meV as determined from the Korringa
relation \cite{SlichterBook}:

\begin{eqnarray}
\frac{1}{T_{1}T}=\left( \frac{4\pi k_{B}}{\hbar }\right)
A^{2}\overline{n}(E_{F})^{2} \label{Korringa_relation}
\end{eqnarray}

\noindent where $\overline{n}%
(E_{F})=0.014 \text{ states/eV/atom}$ is the DOS at the Fermi level for a $%
d\approx 1.4$ nm metallic tube in the tight-binding approximation \cite%
{DresselhausTubesNew}. The above discussed uniformity of the
homogeneous broadening suggests that the heterodimer spins do not
sense separate metallic and semiconducting tubes as it would be
expected based on the geometry of tubes alone
\cite{DresselhausTubesNew}. This can be explained by charge transfer
in the SWCNTs bundles, which shifts the Fermi level and renders all
tubes metallic.

In summary, we observed C$_{59}$N monomer radicals encapsulated in
SWCNTs. The nanotube cage hinders and makes the molecular rotation
anisotropic. We find a low activation energy for charge transfer to
the tubes. At low temperatures, bound C$_{59}$N-C$_{60}$
heterodimers are observed when mixtures of the two fullerenes are
encapsulated. Electron spin-relaxation of the heterodimer shows an
overall metallic behavior of the tubes. The material is a step
toward the realization of confined linear spin-chains, which might
find application in e.g. quantum information processing.

FS acknowledges the Zolt\'{a}n Magyary programme for support. Work
supported by the Austrian Science Funds (FWF) project Nr. 17345, by
the Deutsche Forschungsgemeinschaft (DFG), by the EU projects
MERG-CT-2005-022103 and BIN2-2001-00580 and by the Hungarian State
Grants (OTKA) No. TS049881, F61733, PF63954, and NK60984.

$^{\ast }$ Corresponding author: ferenc.simon@univie.ac.at


\end{document}